\documentclass[12pt]{article}

\usepackage{a4wide}
\usepackage[pdftex,usenames,dvipsnames]{color}
\usepackage{graphics}
\usepackage{amsfonts}

\newcommand{\nit}{\noindent}

\newcommand{\np}{\newpage}
\newcommand{\dsp}{\displaystyle}
\newcommand{\vs}[1]{\vspace{#1 ex}}
\newcommand{\hs}[1]{\hspace{#1 em}}
\newcommand{\bfr}{\begin{flushright}}
\newcommand{\efr}{\end{flushright}}
\newcommand{\bc}{\begin{center}}
\newcommand{\ec}{\end{center}}
\newcommand{\ben}{\begin{enumerate}}
\newcommand{\een}{\end{enumerate}}

\newcommand{\be}{\begin{equation}}
\newcommand{\ee}{\end{equation}}
\newcommand{\ba}{\begin{array}}
\newcommand{\ea}{\end{array}}

\newcommand{\ct}{\cite}
\newcommand{\bit}{\bibitem}

\newcommand{\ag}{\alpha}
\newcommand{\bg}{\beta}

\newcommand{\del}{\delta}

\newcommand{\ve}{\varepsilon}

\newcommand{\kg}{\kappa}
\newcommand{\lb}{\lambda}
\newcommand{\sg}{\sigma}
\newcommand{\rg}{\rho}

\newcommand{\Gam}{\Gamma}

\newcommand{\Sg}{\Sigma}

\newcommand{\lh}{\left(}
\newcommand{\rh}{\right)}

\newcommand{\nb}{\nabla}

\begin{document}

\bc
{\large \bf Spinning Bodies in General Relativity}
\vs{6}

{\large J.W.\ van Holten} 
\vs{4}

{\large Nikhef, Amsterdam NL}
\vs{4} 

November 17, 2015
\ec
\vs{4}

\nit
{\footnotesize 
{\bf Abstract} \\
A covariant hamiltonian formalism for the dynamics of compact spinning bodies in curved space-time 
in the test-particle limit is described. The construction allows a large class of hamiltonians accounting 
for specific properties and interactions of spinning bodies. The dynamics for a minimal and a specific 
non-minimal hamiltonian is discussed. An independent derivation of the equations of motion from an 
appropriate energy-momentum tensor is provided. It is shown how to derive constants of motion, both 
background-independent and background-dependent ones.}
\vs{5}

\nit
{\bf 1.\ Introduction} 
\vs{1}

\nit
General Relativity is an important tool in modern astrophysics. Among other applications it is indispensible 
in modelling the formation and gravitational interactions of neutron stars and black holes as well as the 
emission of gravitational waves. Such modelling needs to take into account the effects of rotation. In 
particular compact objects like neutron stars and solar-mass black holes can achieve high spin rates. Two 
famous examples are the pulsar in the Crab nebula rotating at a rate of 30 Hz and the observed pulsar in 
the binary system PSR1913+16 rotating at 17 Hz \ct{taylor:2005}, neither of which is extreme: some 
millisecond pulsars have spin rates ten times higher \ct{chakrabarty:2003}. 

The dynamics of spinning compact bodies is therefore an important subject in GR. Much work has been 
done on this topic in the past. Most recent investigations build on earlier work by Mathisson, Papapetrou 
and Dixon describing compact bodies in terms of mass-multipole moments 
\ct{mathisson:1937,papapetrou:1951,dixon:1970}. One can identify the motion of the mass monopole with 
the world-line of a test particle and describe the evolution of the other multipoles with reference to this 
world-line. It is also possible to cast the equations of motion in hamiltonian form, but this is rather 
complicated as the canonical momentum is not proportional to the proper velocity and one needs constraints 
to fix the relations between momentum, proper velocity and spin \ct{pirani:1956}-\ct{costa:2014}. 

It is however possible to follow an alternative route using a test-particle approximation formulated in terms 
of non-canonical kinetic momenta and extended with spin degrees of freedom that live on the world-line of 
the particle \ct{dAmbrosi:2015gsa}. Such models have a straightforward hamiltonian formulation still allowing 
much freedom in the specification of the dynamics by the choice of a hamiltonian. 

\np
\nit
{\bf 2.\ Dynamics of spinning particles}
\vs{1}

\nit
The key to this construction is the choice of particle phase space spanned by the position $\xi^{\mu}(\tau)$, 
the covariant momentum $\pi_{\mu}(\tau)$ and the antisymmetric spin-tensor $\Sg^{\mu\nu}(\tau)$. Here 
$\tau$ is the proper-time parametrizing the particle world-line. There exists a closed set of model-independent 
Poisson-Dirac brackets for these variables defined by the fundamental brackets \ct{dAmbrosi:2015gsa}
\be
\ba{l}
\dsp{ \left\{ \xi^{\mu}, \pi_{\nu} \right\} = \del_{\nu}^{\mu}, \hs{2} 
\left\{ \pi_{\mu}, \pi_{\nu} \right\} = \frac{1}{2}\, \Sg^{\kg\lb} R_{\kg\lb\mu\nu}, }\\
 \\
\dsp{ \left\{ \Sg^{\mu\nu}, \pi_{\lb} \right\} = \Gam_{\lb\kg}^{\;\;\;\mu}\, \Sg^{\nu\kg}  
 - \Gam_{\lb\kg}^{\;\;\;\nu}\, \Sg^{\mu\kg}, }\\
 \\
\dsp{ \left\{ \Sg^{\mu\nu}, \Sg^{\kg\lb} \right\} =  g^{\mu\kg} \Sg^{\nu\lb} - g^{\mu\lb} \Sg^{\nu\kg}
 - g^{\nu\kg} \Sg^{\mu\lb} + g^{\nu\lb} \Sg^{\mu\kg}. }
\ea
\label{2.1}
\ee
Remarkably the structure functions of these brackets are the quantities characterizing the geometry of the
space-time manifold on which the particle moves: the metric $g_{\mu\nu}$, the connection coefficients 
$\Gam_{\mu\nu}^{\;\;\;\lb}$ and the Riemann tensor $R_{\kg\lb\mu\nu}$. Their properties guarantee the 
closure of the bracket algebra required by the Jacobi identities for triple brackets. 

To get a complete specification of the dynamics of the spinning particles the brackets must be supplemented
by a proper-time hamiltonian; the minimal choice is the kinetic hamiltonian
\be
H_0 = \frac{1}{2m}\, g^{\mu\nu} \pi_{\mu} \pi_{\nu}, 
\label{2.2}
\ee
where $m$ is the particle mass. Other choices are possible and a relevant example will be discussed later on. 
The equation of motion for any phase-space function $F(x,\pi,\Sg)$ is obtained by computing its bracket with 
the hamiltonian:
\be
\dot{F} = \frac{dF}{d\tau} = \left\{ F, H_0 \right\}.
\label{2.3}
\ee
Defining the proper 4-velocity $u^{\mu} = \dot{\xi}^{\mu}$ the equations of motion for the fundamental 
dynamical degrees of freedom become
\be
\ba{l}
\dsp{ \pi_{\mu} = m g_{\mu\nu} u^{\nu}, \hs{2} 
\frac{Du^{\mu}}{D\tau} = \dot{u}^{\mu} + \Gam_{\lb\nu}^{\;\;\;\mu} u^{\lb} u^{\nu} 
 = \frac{1}{2m}\, \Sg^{\kg\lb} R_{\kg\lb\;\,\nu}^{\;\;\;\,\mu} u^{\nu}, }
\ea
\label{2.4}
\ee
and
\be
\frac{D\Sg^{\mu\nu}}{D\tau} = \dot{\Sg}^{\mu\nu} + u^{\lb} \Gam_{\lb\kg}^{\;\;\;\mu} \Sg^{\kg\nu} 
 + u^{\lb} \Gam_{\lb\kg}^{\;\;\;\nu} \Sg^{\mu\kg} = 0.
\label{2.5}
\ee
The equations show that for a free particle --as defined by the minimal kinetic hamiltonian-- the covariant 
momentum equals the kinetic momentum and that the spin-tensor is covariantly constant. However in the 
presence of a spin- and curvature-dependent force the world-line is not a geodesic and the four-velocity is
not transported parallel to itself. 

Note that the spin-tensor can be decomposed into two space-like vectors $S_{\mu}$ and $Z^{\mu}$:
\be
\Sg^{\mu\nu} = - \frac{1}{\sqrt{-g}}\, \ve^{\mu\nu\kg\lb} u_{\kg} S_{\lb} + u^{\mu} Z^{\nu} - u^{\nu} Z^{\mu}, 
 \hs{2} S^{\mu} u_{\mu} = Z^{\mu} u_{\mu} = 0,
\label{2.6}
\ee
where 
\be
S_{\mu} = \frac{1}{2}\, \sqrt{-g}\, \ve_{\mu\nu\kg\lb} u^{\nu} \Sg^{\kg\lb}, \hs{2}
Z^{\mu} = \Sg^{\mu\nu} u_{\nu}.
\label{2.7}
\ee
Thus $S_{\mu}$ represents the spin proper, reducing in the rest frame to the magnetic components of the 
spin-tensor, whilst $Z^{\mu}$ represents a mass-dipole reducing to the electric components. These 
components themselves are not covariantly constant but their variations are of higher order in the spin-tensor:
\be
D_{\tau} S_{\mu} = \frac{1}{4m}\, \sqrt{-g} \ve_{\mu\nu\kg\lb} R^{\nu}_{\;\rg\sg\tau} u^{\rg} 
 \Sg^{\kg\lb} \Sg^{\sg\tau}, \hs{1}
D_{\tau} Z^{\mu} = \frac{1}{2m}\, R_{\nu\rg\sg\tau} u^{\rg} \Sg^{\mu\nu} \Sg^{\sg\tau}.
\label{2.8}
\ee
Therefore it is not possible to require the permanent vanishing of the mass dipole on the world-line, 
like it is required in the canonical approach by the Pirani condition \ct{pirani:1956}. 
\vs{2}

\nit
{\bf 3.\ Einstein equations} 
\vs{1}

\nit
The equations of motion (\ref{2.4}) and (\ref{2.5}) have been derived from a covariant hamiltonian 
phase-space formulation. However, the same equations can be derived from the Einstein equations 
for a spinning point particle with an appropriate choice of energy-momentum tensor as we now show. 
The issue is that the covariant divergence of the Einstein tensor vanishes because of the Bianchi identities:
\be
\nb^{\mu} G_{\mu\nu} = \nb^{\mu} R_{\mu\nu} - \frac{1}{2}\, \nb_{\nu} R = 0.
\label{3.1}
\ee
Therefore the energy-momentum tensor which provides the source term of the Einstein equations
must have the same property
\be
\nb^{\mu} T_{\mu\nu} = 0.
\label{3.2}
\ee
The energy-momentum tensor of a free spinless particle of mass $m$ moving on a world-line 
$X^{\mu}(\tau)$ is given by the proper-time integral \ct{weinberg:1972}
\be
T_0^{\mu\nu} = \frac{m}{\sqrt{-g}}\, \int d\tau u^{\mu} u^{\nu}\, \del^4\lh x - \xi(\tau)\rh.
\label{3.3}
\ee
The square root is included because we take the $\del$-distribution to be a scalar density of weight 1/2 
such that
\[
\int d^4 y\, \del^4(y - x) f(y) = f(x).
\]
It is then straightforward to establish that after a partial integration
\be
\nb_{\mu} T_0^{\mu\nu} = \frac{m}{\sqrt{-g}} \int d\tau\, \frac{Du^{\nu}}{D\tau}\, \del^4\lh x - \xi(\tau) \rh,
\label{3.4}
\ee
which vanishes if the particle moves on a geodesic. The spin-dependent force in eq.\  (\ref{2.4}) 
can be taken into account by adding to the energy-momentum tensor a term 
\be
T_1^{\mu\nu} =  \nb_{\lb} \left[ \frac{1}{2\sqrt{-g}}\, \int d\tau \lh u^{\mu} \Sg^{\nu\lb} + u^{\nu} \Sg^{\mu\lb} \rh
 \del^4\lh x - \xi(\tau) \rh \right]. 
\label{3.5}
\ee
Computing the covariant divergence of this expression now involves commuting two covariant derivatives
which provides a term with a Riemann tensor plus a term which --after partial integration-- becomes the 
covariant derivative of the spin-tensor: 
\be
\ba{lll}
\nb_{\mu} T_1^{\mu\nu} & = & \dsp{  \frac{1}{2\sqrt{-g}}\, R^{\nu}_{\;\mu\kg\lb} \int d\tau\, u^{\mu} \Sg^{\kg\lb}
 \del^4\lh x - \xi(\tau) \rh }\\
 & & \\
 & & \dsp{ +\, \nb_{\lb} \left[ \frac{1}{2\sqrt{-g}} \int d\tau\, \frac{D\Sg^{\nu\lb}}{D\tau}\, \del^4\lh x - \xi(\tau) \rh \right]. }
\ea
\label{3.6}
\ee
Combining the constributions (\ref{3.4}) and (\ref{3.6}) to the divergence of the total energy-momentum
tensor we get 
\be
\nb_{\mu} T^{\mu\nu} = \nb_{\mu} \lh T_0^{\mu\nu} + T_1^{\mu\nu} \rh = 0,
\label{3.7}
\ee
where the first term on the right-hand side of eq.\ (\ref{3.6}) involving an integral over a $\del$-function 
combines with the expression on the right-hand side of (\ref{3.4}) to form the world-line equation of motion 
(\ref{2.4}), whilst the second term on the right-hand side of eq.\ (\ref{3.6}) involving a derivative of a 
delta-function vanishes separately because of the spin-tensor equation of motion (\ref{2.5}). 
Thus the equations of motion (\ref{2.4}) and (\ref{2.5}) are the necessary and sufficient conditions for the 
divergence of the energy-momentum tensor (\ref{3.7}) to vanish. This provides an independent argument 
for the equations of motion of a spinning particle in curved space-time proposed in section 2.
\vs{2}

\nit
{\bf 4.\ Conservation laws}
\vs{1}

\nit
By eq.\ (\ref{2.3}) a constant of motion for a spinning particle is a quantity $J(x,\pi,\Sg)$ of
which the bracket with the hamiltonian vanishes:
\be
\left\{ J, H_0 \right\} = 0.
\label{4.1}
\ee
There are three background-independent constants of motion. First, the hamiltonian itself:
\be
H_0 = - \frac{m}{2},
\label{4.2}
\ee
where $m$ is the mass of the particle; eq.\ (\ref{4.2}) is equivalent to normalizing proper time such that
$u_{\mu} u^{\mu} = -1$. Furthermore there are two constants of motion constructed as quadratic expressions
in the spin:
\be
I = \frac{1}{2}\, g_{\mu\kg} g_{\nu\lb} \Sg^{\mu\nu} \Sg^{\kg\lb} = S_{\mu} S^{\mu} + Z_{\mu} Z^{\mu}, \hs{2}
D = \frac{1}{8} \sqrt{-g} \ve_{\mu\nu\kg\lb} \Sg^{\mu\nu} \Sg^{\kg\lb} = S_{\mu} Z^{\mu}.
\label{4.3}
\ee
Apart from these generic conservation laws there can be other constants of motion determined by
symmetries of the background space-time. More specifically, a quantity $J$ of the form
\be
J = \ag^{\mu} \pi_{\mu} + \frac{1}{2}\, \bg_{\mu\nu} \Sg^{\mu\nu}, 
\label{4.4}
\ee
is a constant of motion provided 
\be
\nb_{\mu} \ag_{\nu} + \nb_{\nu} \ag_{\mu} = 0, \hs{2} \nb_{\lb} \bg_{\mu\nu} = R_{\mu\nu\lb\kg} \ag^{\kg}.
\label{4.5}
\ee
These equations have solutions if there exists a Killing vector with covariant components $\ag_{\mu}$, 
as one can then find $\bg_{\mu\nu}$ in terms of its curl:
\be
\bg_{\mu\nu} = \frac{1}{2} \lh \nb_{\mu} \ag_{\nu} - \nb_{\nu} \ag_{\mu} \rh. 
\label{4.6}
\ee

\nit
{\bf 5.\ Spin-dependent forces} 
\vs{1}

\nit
An important aspect of the brackets (\ref{2.1}) is their closure independent of the choice of hamiltonian. 
The kinetic hamiltonian $H_0$ is the minimal choice, but it can easily be extended with additional 
interactions. In particular one can introduce a spin-dependent gravitational Stern-Gerlach force 
\ct{dAmbrosi:2015gsa,khriplovich:1999}
proportional to the gradient of the curvature by including a spin-spin interaction with coupling
constant $\kg$ in the hamiltonian: 
\be
H = H_0 + H_{SG}, \hs{1} H_{SG} = \frac{\kg}{4}\, R_{\mu\nu\kg\lb} \Sg^{\mu\nu} \Sg^{\kg\lb}. 
\label{5.1}
\ee
In combination with the brackets (\ref{2.1}) this produces the equations of motion
\be
\pi_{\mu} = m g_{\mu\nu} u^{\nu}, \hs{1}
\frac{Du^{\mu}}{D\tau} = \frac{1}{2m}\, \Sg^{\kg\lb} R_{\kg\lb\;\,\nu}^{\;\;\;\,\mu} u^{\nu}
 - \frac{\kg}{4m}\, \Sg^{\rg\sg} \Sg^{\kg\lb} \nb_{\mu} R_{\rg\sg\kg\lb},
\label{5.2}
\ee
and
\be
\frac{D\Sg^{\mu\nu}}{D\tau} = \kg \Sg^{\rg\sg} \lh R_{\rg\sg\;\,\lb}^{\;\;\;\,\mu} \Sg^{\nu\lb}
 - R_{\rg\sg\;\,\lb}^{\;\;\;\,\nu} \Sg^{\mu\lb} \rh. 
\label{5.3}
\ee
Thus the spin-tensor is no longer covariantly constant, but experiences a non-linear spin-dependent 
force itself. 

As for the minimal case the equations of motion derived from this non-minimal hamiltonian can be
obtained equivalently from the Einstein equations with an appropriately extended energy-momentum
tensor. Define 
\be
\ba{lll}
T_{SG}^{\;\mu\nu} & = & \dsp{ \frac{1}{2}\, \nb_{\kg} \nb_{\lb} \int d\tau \lh \Sg^{\mu\lb} \Sg^{\kg\nu} 
 + \Sg^{\nu\lb} \Sg^{\kg\mu} \rh \frac{1}{\sqrt{-g}}\, \del^4\lh x - X \rh }\\
 & & \\
 & & \dsp{ +\, \frac{1}{4} \int d\tau\, \Sg^{\rg\sg} \lh R_{\rg\sg\lb}^{\;\;\;\;\;\,\nu} \Sg^{\lb\mu}
  + R_{\rg\sg\lb}^{\;\;\;\;\;\,\mu} \Sg^{\lb\nu} \rh \frac{1}{\sqrt{-g}}\, \del^4\lh x - X \rh. }
\ea
\label{5.4}
\ee
Again computing the divergence and using the Ricci identity when commuting covariant derivatives 
one gets a term involving a $\del$-function with support on the world-line and a term involving the 
gradient of this $\del$-function: 
\be
\ba{lll}
\nb_{\mu} T_{SG}^{\;\mu\nu} & = & \dsp{ \frac{1}{4} \int d\tau\, \nb^{\nu} R_{\rg\sg\kg\lb}\, \Sg^{\rg\sg} \Sg^{\kg\lb}
 \frac{1}{\sqrt{-g}}\, \del^4\lh x - X \rh }\\
 & & \\
 & & \dsp{ +\, \frac{1}{2}\, \nb_{\lb} \int d\tau\, \Sg^{\rg\sg} \lh R_{\rg\sg\kg}^{\;\;\;\;\;\,\lb} \Sg^{\kg\nu}
  - R_{\rg\sg\kg}^{\;\;\;\;\;\,\nu} \Sg^{\kg\lb} \rh \frac{1}{\sqrt{-g}}\, \del^4\lh x - X \rh. }
\ea
\label{5.5}
\ee
These terms are to be combined with the similar terms from the divergence of $T_0^{\mu\nu}$ 
and $T_1^{\mu\nu}$ in  eqs.\ (\ref{3.4}) and (\ref{3.6}). It follows that 
\be
\nb_{\mu} \lh T_0^{\mu\nu} + T_1^{\mu\nu} + \kg T_{SG}^{\;\mu\nu} \rh = 0 
\label{5.6}
\ee
if and only if the equations of motion (\ref{5.2}) and (\ref{5.3}) are satisfied. 

Remarkably all of  the conservation laws established for the minimal case carry over unchanged to the 
non-minimal extension with Stern-Gerlach interactions. This is true not only for the generic  constants 
of motion $(H, I, D)$ but also for the background-dependent quantities $J$ associated with Killing
vectors. To prove this it suffices to observe that as a consequence of the Bianchi-identities \ct{dAmbrosi:2015gsa}
\be
\left\{ J , H_{SG} \right\} = \kg\, \Sg^{\mu\nu} \Sg^{\rg\sg} \lh - \frac{1}{4}\, \ag^{\lb} \nb_{\lb} R_{\rg\sg\mu\nu}
 + R_{\rg\sg\mu}^{\;\;\;\;\;\,\lb} \bg_{\lb\nu} \rh = 0.
\label{5.7}
\ee
This holds for all values of the coupling constant $\kg$. 
\vs{2}

\nit
{\bf 6.\ Final remarks} 
\vs{1}

\nit
In this work a relativistic spinning-particle dynamics without constraints has been formulated. 
It was first derived from a purely hamiltonian phase-space approach \ct{dAmbrosi:2015gsa}, but 
an alternative derivation based on the Einstein equations with a suitable energy-momentum tensor 
has been presented here. It is of some interest to observe, that the spin equation of motion 
(\ref{2.5}), (\ref{5.3}) also implies the vanishing divergence of another 3-index tensor 
\be
\ba{lll}
M^{\mu\nu\lb} & = & - M^{\mu\lb\nu} = M_1^{\mu\nu\lb} + \kg M_{SG}^{\mu\nu\lb} \\
 & & \\
 & = & \dsp{ \int d\tau\,  u^{\mu} \Sg^{\nu\lb} \frac{1}{\sqrt{-g}} \del^4(x - \xi)
  + 2 \kg \nb_{\kg} \int d\tau\, \Sg^{\mu\kg} \Sg^{\nu\lb} \frac{1}{\sqrt{-g}} \del^4(x - \xi), }
\ea
\label{6.1}
\ee
Indeed 
\be
\nb_{\mu} M^{\mu\nu\lb} = 
 \int d\tau \left[ \frac{D\Sg^{\nu\lb}}{D\tau} - \kg \Sg^{\rg\sg} \lh R_{\rg\sg\;\,\kg}^{\;\;\;\,\nu} \Sg^{\lb\kg}
  - R_{\rg\sg\;\,\kg}^{\;\;\;\,\lb} \Sg^{\nu\kg} \rh \right] \frac{1}{\sqrt{-g}} \del^4(x - \xi) = 0.
\label{6.2}
\ee
Obviously this tensor is closely related in structure to the usual orbital angular momentum tensor 
in General Relativity. 

A constraint-free spinning-particle dynamics is quite convenient for the analysis of relativistic 
dynamical systems. We have applied it to some astrophysical problems of practical interest, 
like the motion of compact spinning bodies in black-hole space-times. Results will be discussed 
elsewehere \ct{dAmbrosi:2015-4p}. The construction of the corresponding energy-momentum
tensors and the corresponding Einstein equations will also enable the calculation of gravitaional 
waves emitted by spinning bodies, and the evaluation of radiation reaction effects 
\ct{mino:1997,quinn:1997,poisson:2004}. These applications require further investigations. 
\vs{3}

\nit
{\bf Acknowledgement} \\
The author is indebted to G.\ d'Ambrosi and S.\ Satish Kumar for useful discussions and collaboration. 
The work reported is part of the research progamme {\em Gravitational Physics} of the Foundation for
Fundamental Research of Matter (FOM).

\end{document}